\documentclass{aa}  

\usepackage{graphicx}
\usepackage{txfonts}

\usepackage{dsfont} \usepackage{mathtools}

\usepackage{tikz}
\usepackage{graphicx}
\usepackage{tikz-3dplot}
\usetikzlibrary{positioning}
\usetikzlibrary{calc}
\usetikzlibrary{backgrounds}
\usetikzlibrary{shadings}
\usetikzlibrary{decorations.markings, decorations.pathreplacing}

\pgfdeclarelayer{foreground}
\pgfdeclarelayer{background}
\pgfsetlayers{background,main,foreground}

\usepackage{hyperref} \providecommand{\abs}[1]{\lvert#1\rvert}

\providecommand{\norm}[1]{\left\Vert#1\right\Vert}

 \newcommand{\td}{\,\mathrm{d}}

\usepackage{xspace}

\newcommand{\cronos}{\textsc{Cronos}\xspace}
\newcommand{\hll}{\textsc{HLL}\xspace}
\newcommand{\hllc}{\textsc{HLLC}\xspace}

\newcommand{\glorentz}{{u^0}}

\newcommand{\conserved}{{U}}
\newcommand{\fluxes}{{F}}
\newcommand{\fluxesOneD}{{f}}
\newcommand{\primitive}{{w}}
\newcommand{\source}{{S}}
\newcommand{\primitiveM}{\overline{\primitive}}
\newcommand{\conservedM}{\overline{\conserved}}
\newcommand{\fluxesM}{\overline{\fluxes}}
\newcommand{\sourceM}{\overline{\source}}

\newcommand{\vV}{\primitive}

\usepackage[capitalise]{cleveref}
\Crefname{equation}{Eq.}{Eqs.}
\Crefname{figure}{Fig.}{Figs.}
\Crefname{tabular}{Tab.}{Tabs.}
\Crefname{section}{Sect.}{Sects.}  
\begin{document}

\title{Special Relativistic Hydrodynamics with CRONOS}

\author{
  D. Huber \inst{1}
  \and
  R. Kissmann\inst{2}
}

\institute{
  Institut f\"ur Astro- und Teilchenphysik \\
  University of Innsbruck\\
  6020 Innsbruck, Austria\\
  \email{contact@david-huber.eu}
  \and
  Institut f\"ur Astro- und Teilchenphysik \\
  University of Innsbruck\\
  6020 Innsbruck, Austria\\
  \email{ralf.kissmann@uibk.ac.at}
}

\date{Received --; accepted --}

\abstract{We describe the special relativistic extension of the \cronos code, which has been used for studies of gamma-ray binaries in recent years.
The code was designed to be easily adaptable, allowing the user to easily change existing functionalities or introduce new modules tailored to the problem at hand.
Numerically, the equations are treated using a finite-volume Godunov scheme on rectangular grids, which currently support Cartesian, spherical, and cylindrical coordinates.
The employed reconstruction technique, the approximate Riemann solver and the equation of state can be chosen dynamically by the user.
Further, the code was designed with stability and robustness in mind, detecting and mitigating possible failures early on.
We demonstrate the code's capabilities on an extensive set of validation problems.}


\maketitle

\section{Introduction} \label{sec:intro}
Relativistic hydrodynamics is one of the most successful tools to describe complex dynamical phenomena ranging from cosmological scales to those of colliding particles.
General relativity is thereby a necessary ingredient for the understanding of systems with strong gravitational fields, such as accretion discs and the launch of jets in active galactic nuclei \citep[e.g.][]{
Moscibrodzka2014A&A...570A...7M, 
Moscibrodzka2016A&A...586A..38M, 
Komissarov2021NewAR..9201610K} or the mergers of compact objects producing gravitational waves \citep[e.g.][]{Siegel2018ApJ...858...52S}.
Many other phenomena, however, do not involve strongly bent spacetime but still flow velocities near the speed of light, for example 
gamma-ray bursts \citep[e.g.][]{Willingale2017SSRv..207...63W, Zhang2018pgrb.book.....Z}, 
systems with pulsar winds \citep[e.g.][]{Barkov2019MNRAS.485.2041B}, such as gamma-ray binaries \citep[][]{Huber2020, Huber2021} 
and pulsar wind nebulae \citep[e.g.][]{Olmi2019MNRAS.484.5755O} 
or interacting heavy-ion beams in collider experiments \citep[e.g.][]{Busza2018ARNPS..68..339B}.
Their dynamics can therefore be more suitably described in special relativistic frameworks, which we focus on in this work.

Due to the high complexity of many problems in current research, analytical approaches are no longer feasible to explain observed data.
Numerical approaches \citep[see][for a review]{Marti2003LRR.....6....7M, Marti2015LRCA....1....3M} are therefore essential to further our understanding of the related systems.
For this, many modern codes have been presented that treat the equations of special relativistic hydrodynamics (SRHD) and/or magnetohydrodynamics (SRMHD), such as 
\textsc{Pluto} \citep{Mignone2007ApJS..170..228M,Mignone2012ApJS..198....7M},
\textsc{Ramses} \citep{Teyssier2002A&A...385..337T, Lamberts2013A&A...560A..79L},
\textsc{Athena++} \citep{Stone2008ApJS..178..137S, White2016ApJS..225...22W},
\textsc{Amrvac} \citep{Meliani2007MNRAS.376.1189M}, and others.

In this paper, we present the special relativistic extension of \cronos \citep{Kissmann2018ApJS..236...53K}, which was used in comprehensive models for the emission of gamma-ray binaries \citep[see][]{Huber2020, Huber2021}.
\cronos is a finite-volume, Godunov-type code that was developed to solve hyperbolic systems of partial differential equations with astrophysical applications in mind.
A key feature of \cronos consists in its easy adaptability and extensibility, providing a framework with several interfaces for customisations by the user that allows for straightforward incorporation of additional physics, such as 
the transport of energetic particles \citep[see e.g.][]{Reitberger2017ApJ...847...40R}, 
radiative cooling of the plasma \citep[see e.g.][]{Scherer2020MNRAS.493.4172S}, or
the use of more general reference-frames \citep[see e.g.][]{Huber2020}.
Here we describe the extension of \cronos to SRHD, where we did not introduce any new algorithms but select and combine different techniques that have been proven successful in literature.

The paper is structured as follows: in \cref{sec: equations} we introduce the governing equations; the notations and general structure of the code are defined in \cref{sec: notation}; in \cref{sec: numerical scheme} we describe the numerical scheme implemented in \cronos; and validate it on an extensive set on numerical tests in \cref{sec: validation}.
Finally, we comment on the parallel performance of the code in \cref{sec: performance} and give a summary in \cref{sec: summary}. 

\section{System of Equations}\label{sec: equations}

The \cronos code \citep{Kissmann2018ApJS..236...53K} was in general developed to solve a set of hyperbolic conservation laws of the form
\begin{equation}
  \partial_t \conserved + \nabla \cdot \mathbf{\fluxes} \left( \conserved \right) = \source,
  \label{eq: conservation equation}
\end{equation}
where $\conserved$ denotes a set of conserved quantities, $\mathbf{\fluxes}$ the respective fluxes, and $\source$ an additional source term.

\subsection{Special relativistic hydrodynamics}

In analogy to classical hydrodynamics, the dynamics of a relativistic fluid can be expressed as such a system \citep{Landau1959flme.book.....L}.
The respective equations for the conservation of rest-mass, energy and momentum read as
\begin{subequations}
  \label{eq: srhd system}
\begin{eqnarray}
  \partial_t D + \nabla \cdot \left( D \mathbf{v}  \right)= S_D,
  \label{eq: mass cons}\\
  \partial_t \tau + \nabla \cdot \left( (\tau + p) \, \mathbf{v} \right) = S_\tau,
  \label{eq: energy cons}\\
  \partial_t \mathbf{m} + \nabla \cdot \left( \mathbf{m} \otimes \mathbf{v} + p \, \mathbf{\mathds{1}} \right)=\mathbf{S}_m,
  \label{eq: momentum cons}
\end{eqnarray}
\end{subequations}
where $\conserved = (D, \tau, m^j)$ correspond to the conserved variables, i.e. the mass, energy, and momentum density in the $j$-th coordinate direction, respectively, and $\source = (S_D, S_\tau, S_{m^j})$ to their respectively source terms.
The fluxes further involve the fluid's three-velocity $\mathbf{v}$ and pressure $p$.
All quantities are given in the laboratory frame and in natural units, i.e. with the speed of light $c=1$.

It should be noted that the momentum fluxes in cref{eq: momentum cons} are given by the rank-2 stress tensor.
Additional geometrical source terms might therefore arise implicitly when evaluating the divergence in non-Cartesian coordinates (see cref{sec: coordinate systems}).

Other codes treat the conservation of energy by solving an equivalent equation for $E = \tau + D$ \citep[see e.g.][]{Mignone2007ApJS..170..228M, Lamberts2013A&A...560A..79L}.
In the low pressure, non-relativistic regime, the variables $D$ and $E$, however, can take very similar values, which can lead to severe truncation errors.
This can be avoided by evolving the variable $\tau = E - D$, instead, which we therefore also use in \cronos{}.
This choice is solely motivated by numerical reasons and is therefore used in many codes \citep[see e.g.][]{Banyuls1997ApJ...476..221B, Aloy1999ApJS..122..151A, Baiotti2005PhRvD..71b4035B, Meliani2007MNRAS.376.1189M, Rezzolla2013}.

The conserved variables and the corresponding fluxes are further expressed in terms of the primitive fluid variables $\primitive = (\rho, p, u^i)$, that are the fluid's mass density $\rho$, pressure $p$, and the spatial part of the four-velocity vector $u^i$, respectively.
The relations read as
\begin{equation}
  \begin{split}
  \conserved \left( \primitive \right) &=
  (D, \tau, m^j) 
  \\ &=
\left( \rho \glorentz, \rho \glorentz (h \glorentz - 1) - p, \rho h  \glorentz u^j \right),
  \end{split}
\end{equation}
where the fluid's bulk Lorentz factor is denoted by $\glorentz$ and its specific enthalpy by $h$.

In contrast to other implementations that employ the three-velocity \citep[e.g.][]{Mignone2007ApJS..170..228M, Lamberts2013A&A...560A..79L} as primitive quantity, we use the four-velocity, instead, and infer the former via $\mathbf{v} = \mathbf{u} / \glorentz$.
This choice is more natural for numerical purposes since $\abs{\mathbf{u}}$ is not bound by the speed of light -- a constraint that one would have to guarantee for $\abs{\mathbf{v}}$ at all times.
The computation of the Lorentz factor via $(\glorentz)^2 = 1 + \abs{\mathbf{u}}^2$ is also more robust by naturally avoiding truncation errors as compared to $(\glorentz)^2 = (1 - \abs{\mathbf{v}}^2)^{-1}$.

The presented system of five equations in six unknowns is closed by specifying an additional equation of state (EoS), which is usually provided in the form of $h = h(\Theta)$ with $\Theta = \frac{p}{\rho}$.
In \cronos we consider two EoS: The ideal one, assuming a constant adiabatic index $\Gamma$ and the Taub-Mathews (TM) EoS \citep{Mathews1971ApJ...165..147M}, which approaches $\Gamma \rightarrow \frac{4}{3}$ in the high-temperature limit, and $\Gamma \rightarrow \frac{5}{3}$ in the low-temperature limit.
The specific enthalpies are given by
\begin{equation}
  h(\Theta) =
  \begin{cases}
    1 + \frac{\Gamma}{\Gamma - 1} \Theta & \mathrm{ideal \, EoS}\\
    \frac{5}{2} \Theta + \sqrt{\frac 9 4 \Theta^2 + 1} & \mathrm{TM \, EoS}
  \end{cases}
  ,
  \label{eq: eos enthalpies}
\end{equation}
respectively.
For our current purposes, these EoSs are sufficient, although it should be noted that there are many more available in the literature \citep[see e.g.][]{Synge1957, Rezzolla2013}.
Due to the modular design of \cronos, additional EoSs can be easily added in the future by supplying the relevant routines.

\subsection{Passive tracer fields}\label{sec: tracer}
\cronos provides the option to solve user-defined conservation laws alongside the ones governing the fluid dynamics \citep[see also][]{Kissmann2018ApJS..236...53K}.
A prominent use case for this is the treatment of tracers fields $\psi$, which are passively advected with the fluid flow.
Their value remains constant along fluid-lines, following the transport equation
\begin{equation}
  \left(\partial_t + \mathbf{v} \cdot \nabla \right) \psi= 0 \label{eq: tracer fluid lines}
  .
\end{equation}
A corresponding conservation equation for $\Psi = \psi D$ can be found by combining cref{eq: mass cons} and cref{eq: tracer fluid lines}, yielding
\begin{equation}
  \partial_t \Psi + \nabla \cdot \left( \Psi \mathbf{v} \right) = 0 \label{eq: tracer conservation}
  .
\end{equation}
This equation is treated in analogy to the mass-conservation equation cref{eq: mass cons} by the fluid solvers. 

\section{Notation and General Structure} \label{sec: notation}

\subsection{Coordinate systems}\label{sec: coordinate systems}
In \cronos we employ orthogonal, curvilinear coordinate systems in 3 dimensions, which are expressed in the generic variables $(x^1, x^2, x^3)$ and the corresponding spatial metric tensor $\gamma = \text{diag}(h^2_1, h^2_2, h^2_3)$ given by the scaling factors $h_i$.
Specifically, we have implemented Cartesian, cylindrical and spherical coordinates (see also \cref{tab: coordinates}).
For the latter two, the divergence of the rank 2 stress-tensor in the momentum-conservation equation \cref{eq: momentum cons} yields non-zero geometrical source terms.
The expression can be expanded component-wise in terms of the standard vector-divergence in the respective coordinate system, yielding
\begin{equation}
  \partial_t m^j + \nabla \cdot \left( m^j \mathbf{v} + p \, \mathbf{e}_j \right)= S_{m^j} + G_{m^j}
  \label{eq: momentum eq with geom source}
\end{equation}
where $\mathbf{e}_j$ denotes the unit base vector along the $j$-th coordinate direction (see \cref{app: geom source terms}).

\subsection{Normalisation}
For the numerical treatment of the equations, \cronos internally uses normalised quantities in the form $X = X_0 \hat{X}$, where $X$ is a physical quantity, $X_0$ a normalisation constant and $\hat{X}$ the resulting unit-free, normalised quantity that is treated by the solvers \citep{Kissmann2018ApJS..236...53K}.
Our implementation foresees four base quantities that can be normalised independently, that correspond to the length, time, mass and number density.
The latter is not strictly necessary to have a complete system of units.
However, it has been proven to be useful in practice since employing a normalisation derived from the length-scale can lead to unwieldy large values for the number-density.

The normalisations can be chosen freely by the user depending on the problem at hand by specifying the normalisation constants for four different quantities, which can also include derived quantities (e.g. energy density, velocity, etc.).
If constants for derived quantities are supplied, the normalisations for the non-specified base quantities are obtained automatically via physical relations.
If the specified quantities form an inconsistent system of units, the procedure is aborted and the user is notified.
In contrast to classical hydrodynamics, velocities are always normalised to the speed of light in the relativistic case, as implicitly imposed by \cref{eq: srhd system}.
The user is hence only allowed to specify the three remaining normalisations to complete the system.

Finally, the normalised equations, corresponding to \cref{eq: srhd system}, that are solved by \cronos are
\begin{subequations}
  \label{eq: srhd system normalised}
  \begin{eqnarray}
    \partial_{\hat{t}} \hat{D} + \hat{\nabla} \cdot \left( \hat{D} \, {\hat{\mathbf{v}}}  \right)= \hat{S}_D
    \label{eq: numerical system D}
    \\
    \partial_{\hat{t}} \hat{\tau} + \hat{\nabla} \cdot \left((\hat{\tau} + \hat{p}) \, \hat{\mathbf{v}}  \right)= \hat{S}_\tau
    \label{eq: numerical system tau}
    \\
    \partial_{\hat{t}} {\hat{m}}^j + \hat{\nabla} \cdot \left( {\hat{m}}^j {\hat{\mathbf{v}}} + \hat{p} \, \hat{\mathbf{e}}_j \right)=  \hat{S}_{m^j} +  \hat{G}_{m^j}
    .
    \label{eq: numerical system m}
  \end{eqnarray}
\end{subequations}
For the rest of this work, we will exclusively use normalised quantities and omit the hat sign on these for better readability.
Once a solution in normalised quantities was obtained, physical quantities can be easily computed in a post-processing step using the corresponding normalisation constants, which are therefore stored together with the data in the output files.

\subsection{Grid structure}
To solve the set of equations \cref{eq: srhd system normalised}, \cronos employs a finite-volume scheme, decomposing the computational volume into a set of disjoint cells \citep[see e.g.][]{Leveque2002}.
While the shapes of the cells are in principle irrelevant for such a scheme, we restrict ourselves to three-dimensional, rectangular grids for practical reasons.

The division into cells is achieved by subdividing each coordinate axis $d \in \lbrace 1,2,3 \rbrace$ spanning the computational domain $(x^d_b, x^d_e)$ into $N^d$ intervals $\omega^d_{i^d}$, which are addressed by the index $i^d \in \lbrace 0, \cdots , N^d - 1 \rbrace$.
The edges of the individual intervals $\omega^d_{i^d} = \left(x^d_{i^d-1/2}, x^d_{i^d+1/2}\right)$ can be freely chosen as long as they are strictly increasing and first and last edges match the beginning and the end of the computational domain, i.e. $x^d_{-1/2} = x^d_b$ and $x^d_{N^d-1/2} = x^d_e$, respectively.
Usually, the edges are chosen to be spaced linearly, however, also non-linear options are implemented \citep[see][]{Kissmann2018ApJS..236...53K}.

The individual cells are then given by
$C_I = \omega^1_{i^1} \times \omega^2_{i^2} \times \omega^3_{i^3}$
and are correspondingly addressed by the index triplet
$I =(i^1, i^2, i^3)$.
The centre of a cell $x^d_I \coloneqq (x^d_{i^d-1/2} + x^d_{i^d+1/2}) / 2$ is defined to be located at the midpoint of each coordinate interval.

A cell $C_I$ borders on a set of neighbouring cells with indices $J \in N(I)$.
The interfaces to its neighbours $\Sigma_{I,J}$ can therefore be used to decompose the cells surface as $\partial C_I = {\underset{J \in N(I)}{\dot{\bigcup}}} \Sigma_{I,J}$.
For rectangular grids, the cell interface normals $\mathbf{n}_{I,J}$ are further (anti-)parallel to the $d_{I,J}\,$-th coordinate basis vectors yielding $\mathbf{n}_{I,J} \cdot \mathbf{e}_{d_{I,J}} =: \sigma_{I,J} = \pm 1$

In contrast to finite difference schemes, where the values of dynamical variables are stored as point-values on a grid, in a finite-volume approach, cell-averaged values are stored.
The averaged primitive variables $\bar{\primitive}$, which are initially set by the user, are defined by
\begin{equation}
  \bar{\primitive}_I \coloneqq \frac{1}{\Delta V_I} \int_{C_I} \primitive \td V \quad \text{with} \quad \Delta V_I \coloneqq \int_{C_I} \td V.
\end{equation}
The corresponding set of averaged conserved variables $\bar{\conserved}$ is defined in analogy.

\subsection{Semi-discrete finite-volume scheme}\label{sec: finite volume}
To obtain a numerical scheme to update the cell averages $\primitiveM$ or respectively $\conservedM$, the set of conservation equations \cref{eq: conservation equation} is integrated over the extent of a cell $C_I$.
Substituting the respective cell-averages and applying Gauss's theorem yields
\begin{align}
  \frac{\td}{\td t} \conservedM_I = \sourceM_I - \frac{1}{\Delta V_I}\int_{\partial C_I} \mathbf{\fluxes} \cdot \td \mathbf{A}
  \label{eq: FV 3}
\end{align}
where $\sourceM$ denotes the cell averaged source term.
Splitting the cell's surface into disjoint interfaces to its neighbours and using the interface-averaged fluxes defined as
\begin{equation}
  \fluxesM_{I,J} \coloneqq \frac{\sigma_{I,J}}{\Delta A_{I,J}} \int_{\Sigma_{I,J}}
F^{d_{I,J}}
  \td A, \; \; \Delta A_{I,J} \coloneqq \int_{\Sigma_{I,J}} \td A
  \label{eq: flux interface average}
\end{equation}
allows \cref{eq: FV 3} to be rewritten as
\begin{equation}
  \frac{\td}{\td t} \conservedM_I = \bar{\mathcal{L}}_I \coloneqq
  \sourceM_I - \frac{1}{\Delta V_I} \sum_{J \in N(I)} \Delta A_{I,J}  \,  \fluxesM_{I,J}
  \label{eq: semi-discrete finite volume formulation}
\end{equation}
With this, the initial system of partial differential equations in \cref{eq: conservation equation} has been cast into a system of ordinary differential equations (ODEs) for each grid cell.
This is usually referred to as a semi-discrete scheme, since the problem was discretised in space but not in time, having the advantage that the spatial and temporal parts of the problem are conceptually decoupled \citep[see e.g. also][]{Kurganov2000JCoPh.160..241K, Leveque2002}.

\section{Numerical scheme} \label{sec: numerical scheme}

In \cronos we follow the general idea presented by \citet{Godunov1959} but use higher-order schemes in both space and time, instead.
First, we obtain interface-centred point-values of the fluid variables at both sides of a given cell interface using a second-order spatial reconstruction (see \cref{sec: reconstruction}).
The resulting jump in the reconstructed quantities can be viewed as a local, one-dimensional Riemann problem, which is solved by approximate Riemann solvers (see \cref{sec: riemann solver}) to obtain an estimation for the fluxes $\fluxesM_{I,J}$ in \cref{eq: semi-discrete finite volume formulation}.
Finally, advancing the solution in time is handled by a second-order Runge-Kutta scheme (see \cref{sec: time integration}).

\subsection{Reconstruction}\label{sec: reconstruction}
To provide the initial data for the Riemann problems to be solved at the cell interfaces, a spatial reconstruction is used.
The order of the reconstruction thereby critically determines the spatial accuracy of the overall method.
In \cronos we apply piecewise-linear reconstructions for every coordinate direction independently, using minmod \citep[see][]{VanLeer1979JCoPh..32..101V, Harten1983} and van-Leer \citep[see][]{VanLeer1977JCoPh..23..276V} slope limiters.
The reconstruction methods employed for special relativistic hydrodynamics are identical to the ones for classical hydrodynamics \citep[see][]{Kissmann2018ApJS..236...53K}.

The reconstruction is applied for primitive variables, yielding the left- (L) and right-handed (R) states $\primitive^L_{I,J}$ and $\primitive^R_{I,J}$, respectively, at a given cell interface $\Sigma_{I,J}$.
Therefrom, the corresponding point-values of the conserved variables $\conserved^{L/R}_{I,J}$ are computed subsequently.
Since the flux function in \cref{eq: srhd system} is expressed in terms of primitive variables, this approach has the additional advantage that the more expensive variable conversion from conserved to primitive ones (see \cref{sec: inversion}) can be avoided.

As opposed to Newtonian hydrodynamics, where the fluid speed is generally not bound, it has to remain subluminal at all times for relativistic hydrodynamics.
This constraint can be easily violated if three-velocity components are reconstructed independently from each other.
In our case, related problems are avoided from the outset since we use and reconstruct spatial four-velocity components instead, which are not subject to such a constraint.

\subsection{Approximate Riemann solvers} \label{sec: riemann solver}
For a given cell interface $\Sigma_{I,J}$, the reconstructed states at both sides are used to define a local Riemann problem that can be formalised as
\begin{equation}
  \primitive(t_0, \xi) =
  \begin{cases}
    \primitive^L_{I,J} & \xi < \xi_{I,J} \\
    \primitive^R_{I,J} & \xi > \xi_{I,J}
  \end{cases},
  \label{eq: riemann problem state}
\end{equation}
with $\xi = x^{d_{I,J}}$ denoting the coordinate in the direction of the interface normal.
Assuming the cell-interface to be flat, the conservation equations \cref{eq: conservation equation} governing the evolution of the Riemann problem reduce to
\begin{equation}
  \partial_t \conserved + \partial_{\xi} f(\primitive) = 0
  \label{eq: riemann problem conservation}
  ,
\end{equation}
where $f = \mathbf{e}_{d_{I,J}} \cdot \mathbf{\fluxes}= \fluxes^{d_{I,J}}$ corresponds to the flux in the direction of the interface normal.

Although relativistic Riemann problems cannot be solved in a closed analytic form, semi-analytic iterative solvers have been developed that can compute their solution to arbitrary precision \citep[see e.g.][]{Marti1994JFM...258..317M, Pons2000JFM...422..125P}.
Employing such solvers, however, is cumbersome and computationally very expensive, which makes them unsuitable for an application in simulation codes.

Instead, approximate Riemann solvers have been developed making simplifying assumptions on the waves emerging from the initial discontinuity.
This significantly reduces the computational cost, while still yielding good approximations for the fluxes at the cell interfaces.
In \cronos we employ the \hll and the \hllc special-relativistic Riemann solvers \citep{Mignone2005MNRAS.364..126M}.
Since every cell interface is treated individually and identically by the solver, we will omit the cell indices, e.g. $\primitive_{L} \equiv \primitive^{L}_{I,J}$, for the remainder of this section for the sake of better readability.

\subsubsection{HLL solver}
One of the most commonly employed Riemann solvers is the \hll solver first proposed by \citet{Harten1983} for the classical Euler equations.
The generalisation to the relativistic case is straightforward and was for the first time used in \citet{Schneider1993JCoPh.105...92S}.
Here, we will briefly summarise the expressions relevant for the implementation in \cronos.

The solver assumes that the Riemann problem produces two shock waves propagating with speeds $\lambda_L$ and $\lambda_R$ computed from \cref{eq: wave speed estimations}.
Only one constant intermediate state $\conserved^*$ is considered to be formed between both shocks, while the states ahead of the shocks remain unchanged.
The intermediate state can be computed directly from the Rankine-Hugoniot jump conditions at both shocks
\begin{equation}
  \conserved^* = \frac{\lambda_R \conserved_R - \lambda_L \conserved_L + \fluxesOneD_L - \fluxesOneD_R}{\lambda_R - \lambda_L}
  ,
  \label{eq: HLL intermediate state}
\end{equation}
using the fluxes of the undisturbed states at either side $f_{L/R} = f(\primitive_{L/R})$.
The corresponding fluxes in the intermediate region are then given by
\begin{align}
  \fluxesOneD^* & = \frac{\lambda_R \fluxesOneD_L - \lambda_L \fluxesOneD_R + \lambda_L \lambda_R \left( \conserved_R - \conserved_L \right)}{\lambda_R - \lambda_L}
  .
\end{align}
Depending on the signal speeds $\lambda_{L/R}$, the numerical cell-interfaces fluxes are ultimately given by
\begin{equation}
  \fluxes^{\hll} = \begin{cases}
    \fluxesOneD_L & \text{for}  \quad 0 < \lambda_L             \\
    \fluxesOneD^* & \text{for}  \quad \lambda_L < 0 < \lambda_R \\
    \fluxesOneD_R & \text{for}  \quad \lambda_R < 0.
  \end{cases} \label{eq: HLL fluxes}
\end{equation}
Although the \hll solver is quite simple and computationally inexpensive, it yields good and robust approximations for the cell-interface fluxes.
In general, the solver has been found to perform well at shock and rarefaction waves.
However, since contact discontinuities were not considered in the simplified wavefan, the \hll solver introduces additional numerical dissipation at such.

\subsubsection{HLLC solver}
A natural improvement to the \hll solver is to restore the neglected contact discontinuity.
The corresponding approximate Riemann solver, which considers all physical intermediate states, is known as \hllc solver and was first presented by \citet{Toro1997} for the classical Euler equations and later extended to special relativistic hydrodynamics by \citet{Mignone2005MNRAS.364..126M}.
Again, we will briefly summarise the quantities relevant for our implementation in \cronos.

The solver assumes the emerging wavefan to be composed of two shocks together with a contact discontinuity in between, separating two intermediate states $\conserved_s^*$ with $s\in \lbrace L,R \rbrace $, respectively.
Using the Rankine-Hugoniot jump conditions for both shock waves and assuming that the intermediate-state fluxes take the following form $f^*_s = f(\primitive_s^*)$, which is not necessarily the case in general \citep{Mignone2005MNRAS.364..126M}, the speed of the contact discontinuity $\lambda^*$ is given by
\begin{multline}
  \left(\lambda^*\right)^2 \left[ \lambda_L A_R - \lambda_R A_L \right]
  + \left[B_R - B_L\right] \\
  - \lambda^* \left[A_R - A_L   + \lambda_L B_R - \lambda_R B_L \right] = 0
  ,
\end{multline}
with
\begin{eqnarray}
  A_s \coloneqq \lambda_s (\tau_s + D_s) - m^x_s \\
  B_s \coloneqq m^x_s ( \lambda_s - v^x_s) - p_s.
\end{eqnarray}
We solve this equation analytically, taking only the root with the minus sign into account since this is the only physically acceptable one \citep{Mignone2005MNRAS.364..126M}.

After computing the intermediate-state pressure for one side
\begin{equation}
  p_s^*= \frac{\lambda^* A_s - B_s}{1 - \lambda_s \lambda^*}
\end{equation}
and using that the pressure is constant over the contact discontinuity $p^* \equiv p_L^* \equiv p_R^*$, the remaining quantities of the intermediate states are given by
\begin{align}
  m^{x*}_s (\lambda_s - \lambda^*)  & = m^{x}_s (\lambda_s - v^x_s) - p_s +p^* \label{eq: RK HLLC 2}                \\
  \tau^*_s (\lambda_s - \lambda^*)  & = \tau_s (\lambda_s - v^x_s) - v^x_s p_s + \lambda^* p^*\label{eq: RK HLLC 4} \\
  \zeta^*_s (\lambda_s - \lambda^*) & = \zeta_s (\lambda_s - v^x_s) \label{eq: RK HLLC 5}
  .
\end{align}
Here, $\zeta$ represents any passively-advected, conserved quantity such as the mass density $D$, the momentum densities $m^{t_1}, m^{t_2}$ tangential to the interface or any generic tracer field $\Psi$.
Finally, the numerical fluxes at the cell interfaces are obtained from
\begin{equation}
  \fluxes^{\hllc} = \begin{cases}
    \fluxesOneD_L   & \text{for}  \quad 0 < \lambda_L             \\
    \fluxesOneD^*_L & \text{for}  \quad \lambda_L < 0 < \lambda^* \\
    \fluxesOneD^*_R & \text{for}  \quad \lambda^* < 0 < \lambda_R \\
    \fluxesOneD_R   & \text{for}  \quad \lambda_R < 0
    .
  \end{cases} \label{eq: HLLC fluxes}
\end{equation}
To avoid a direct evaluation of the flux function $f(\primitive^*_s)$ for the intermediate states, which would require a variable inversion, the intermediate-state numerical fluxes are computed from the Rankine-Hugoniot conditions instead
\begin{equation}
  f_s^* = f_s + \lambda_s \left( U^*_s - U_s \right).
\end{equation}
The \hllc solver has been shown to have superior performance compared to the more simple \hll solver.
However, it has also been reported to potentially suffer from the carbuncle-phenomenon \citep[see e.g.][]{Quirk1994IJNMF..18..555Q,Dumbser2004JCoPh.197..647D}.

\subsubsection{Wave speed estimates}
Both, the \hll and \hllc Riemann solvers require an estimate for the fastest signal speeds to the left and to the right of the discontinuity $\lambda_L, \lambda_R$, respectively \citep[see also][]{Toro1997}.
They can be obtained from the smallest and largest eigenvalues of the system \cref{eq: riemann problem conservation} and were first computed by \citet{Davis1988} for the relativistic case.
Although this is not the only possible estimate \citep[e.g. the Roe average, see][]{Roe1981LNP...141..354R}, it is commonly used in many codes \citep[see e.g.][]{Mignone2007ApJS..170..228M,Lamberts2013A&A...560A..79L}.
In \cronos we use
\begin{subequations}
  \label{eq: wave speed estimations}
  \begin{align}
    \lambda_L & = \min \left( \lambda_- \left( \primitive_L \right), \lambda_- \left( \primitive_R \right) \right) \label{eq: wave speed estimation L}  \\
    \lambda_R & = \max \left( \lambda_+ \left( \primitive_L \right), \lambda_+ \left( \primitive_R \right) \right), \label{eq: wave speed estimation R}
  \end{align}
\end{subequations}
where $\lambda_\pm$ are the smallest and largest eigenvalues of the system, respectively, which take the form
\begin{equation}
  \lambda_\pm (\primitive) = \frac{\glorentz u^d \pm \sqrt{u_s^2 \left( \glorentz^2 - {u^d}^2 + u_s^2 \right)}}{\glorentz^2 + u_s^2}
  \label{eq: characteristic velocities}
  ,
\end{equation}
with $d=d_{I,J}$ cooresponding to the direction of the interface normal, $u_s^2 \coloneqq c_s^2 / (1 - c_s^2)$ and the sound speed \citep[taken from][]{Mignone2007MNRAS.378.1118M}
\begin{equation}
  c^2_s(\Theta) =
  \begin{cases}
    \frac{\Gamma \Theta}{h}                                & \mathrm{ideal \, EoS} \\
    \frac{\Theta}{3 h} + \frac{5 h - 8 \Theta}{h - \Theta} & \mathrm{TM \, EoS},
  \end{cases}
  \label{eq: sound speed}
\end{equation}
with $h(\Theta)$ computed from \cref{eq: eos enthalpies}.

\subsection{Time integration}\label{sec: time integration}
In \cronos, we treat the system of ODEs in time, arising in the semi-discrete framework \cref{eq: semi-discrete finite volume formulation}, using a second-order total-variation-diminishing Runge-Kutta scheme, i.e. Heun's method \citep[see][]{Shu1988}
\begin{subequations}
  \label{eq: heun scheme}
  \begin{align}
    \conservedM^*_I     & = \conservedM^n_I + \bar{\mathcal{L}}_I\left(\primitive^n\right) \Delta t                                                 \\
    \conservedM^{n+1}_I & =  \frac{1}{2} \left(  \conservedM^n_I + \conservedM^*_I +  \bar{\mathcal{L}}_I\left(\primitive^*\right) \Delta t \right)
    ,
  \end{align}
\end{subequations}
where the indices $n$, $n+1$ and $*$ indicate values at the current, the next and an intermediate time-step.

The evaluation of the dimensionally-unsplit operator $\bar{\mathcal{L}}_I$ for a given cell requires the computation of the numerical fluxes at every cell interface.
This involves the solution of Riemann problems, which is the most expensive part of the scheme.
Since $\bar{F}_{I,J} = - \bar{F}_{J,I}$ by construction, we iterate over interfaces in the computational domain instead of over cells to avoid unnecessary double computations.
In practice, we further iterate over one-dimensional pencils for each coordinate, which allows efficient utilisation of the coordinate-wise nature of the reconstructions.
This is followed by another iteration over the whole domain regarding the source term, where we evaluate the respective expressions at the cell centres as
\begin{equation}
  \sourceM_I = S( t, x_I, \primitiveM_I).
\end{equation}

The step size for the numerical time integration is constrained by the CFL-condition \citep[see][]{Courant1928MatAn.100...32C} and is obtained from
\begin{equation}
  \Delta t =  C_\text{CFL} \min_{I,J} \left(\frac{\Delta L^{d(I,J)}_I}{\lambda^\text{max}_{I,J}} \right)
  ,
  \label{eq: CFL}
\end{equation}
where $C_\text{CFL}$ is the Courant number and $\lambda^\text{max}_{I,J}$ the largest absolute value of the signal speeds computed at the cell interface $\Sigma_{I,J}$, which is retrieved from the characteristic speeds that have been computed for the Riemann solvers.
We allow the user to set a specific value for $C_\text{CFL}$ in each parameter file, individually.
For dimensionally unsplit schemes, such as the one described here, the Courant number has to be chosen as $C_\mathrm{CFL} \leq 1/N_d$ to maintain stability in any case \citep[see also][]{Lamberts2013A&A...560A..79L}, where $N_d$ is the dimension of the problem.
Depending on the problem, however, the condition might be eased, e.g. if the time step is determined by a radially symmetric flow, a Courant number of $C_\mathrm{CFL} \leq 1/\sqrt{N_d}$ is sufficient.
The time step is further adapted if an output checkpoint or the end of the simulation is reached.

In contrast to Newtonian hydrodynamics, where the signal speeds $\lambda^\mathrm{max}$ are not limited in general, they are bound by the speed of light in relativity.
In practice, this implies that the step size computed from \cref{eq: CFL} will not become smaller than $\Delta t_\mathrm{min} =  C_\text{CFL} \min_{d, I} \left( \Delta L^d_I \right) / c$, which only depends on the extension of the smallest cell.
Such a lower bound has been proven to be useful in practice for the estimation of maximum simulation runtimes.

After obtaining a solution at $t^{n+1}$, user-supplied routines are called, e.g. to reset the solution in a given region, etc., before the time integration procedure starts again.

\subsection{Variable inversion}\label{sec: inversion}
Since the fluxes are given in terms of the primitive fluid variables $\primitive= \left( \rho, p, u^j \right)$ but the time integration has to be performed on conserved variables $\conserved = \left( D, \tau, m^j \right)$ to obtain a shock-capturing scheme, transformations between the two sets are required.
While the conversion from primitive quantities to conserved ones is trivial and can be implemented in a straightforward manner, the conversion in the other direction is not and involves solving non-linear equations for at least one of the quantities.
For this, we closely followed the approach presented by \citet{Mignone2007MNRAS.378.1118M}, which was initially developed for relativistic MHD, and we adapted for our purposes \citep[similar to][]{Lamberts2013A&A...560A..79L}.
We solve the non-linear equation
\begin{equation}
  f(W') \coloneqq W' - \tau - p(\chi, \rho) = 0, \label{eq: inversion equation}
\end{equation}
where $ W' = D \left( h \glorentz - 1\right)$ and $ \chi = \rho ( h-1)$.
The function $p(\chi, \rho)$ depends on the chosen EoS and can be found in \cref{tab: inversion thermodynamic quantities} for the different cases implemented in \cronos.
To evaluate $f$ for a given $W'$, we first compute the corresponding four-speed via
\begin{equation}
  \abs{\mathbf{u}}^2(W') = \frac{\abs{\mathbf{m}}^2}{(W' + D)^2 - \abs{\mathbf{m}}^2},
\end{equation}
which is used to obtain \citep{Mignone2007MNRAS.378.1118M}
\begin{equation}
  \chi(W', \abs{\mathbf{u}}^2) = \frac{W'}{\glorentz^2} - \frac{D\abs{\mathbf{u}}^2}{\glorentz^2(1 + \glorentz)}
\end{equation}
and
\begin{equation}
  \rho(W', \abs{\mathbf{u}}^2) = \frac D \glorentz \label{eq: inversion rho}
\end{equation}
employing the identity $(\glorentz)^2 = 1 + \abs{\mathbf{u}}^2$.

We solve \cref{eq: inversion equation} numerically using an iterative Newton-Raphson scheme until the relative change is below a certain threshold $\epsilon$ that can be set by the user depending on the problem (typically $\epsilon = 10^{-8}$).
It can be shown analytically that the scheme is accurate in the ultra-relativistic limit as long as $\glorentz \lesssim \epsilon^{-1/2}$ and $p/(\rho \glorentz^2) \gtrsim \epsilon$ \citep[see][]{Mignone2007MNRAS.378.1118M}, which might require an adjustment of $\epsilon$ for certain problems.

The derivative $\td f / \td W'$ required by the algorithm is expressed as
\begin{equation}
  \frac{\td f}{\td W'} = 1 - \frac{\partial p}{\partial \chi} \frac{\td \chi}{\td W'} - \frac{\partial p}{\partial \rho} \frac{\td \rho}{\td W'}
  ,
\end{equation}
where the kinematical derivatives are given by
\begin{eqnarray}
  \frac{\td \chi}{\td W'} = \frac{1}{\glorentz^2} + \glorentz \left( D + 2 \glorentz \chi \right) \frac{\abs{\mathbf{m}}^2}{\left( W' + D \right)^3} \\
  \frac{\td \rho}{\td W'} = \frac{D \glorentz \abs{\mathbf{m}}^2}{\left( W' + D \right)^3}
  .
\end{eqnarray}
The derivatives involving the pressure $\partial_\chi p$ and $\partial_\rho p$ again depend on the chosen EoS and are also listed in \cref{tab: inversion thermodynamic quantities}.

The Newton-Raphson iteration is started using the initial value
\begin{equation}
  W'_{0} = \frac{1}{3} \left( 2 E + \sqrt{4 E^2 - 3 \abs{\mathbf{m}}^2} \right) - D.
\end{equation}
This choice guarantees pressure positivity given the conservative state is physically admissible \citep[][]{Mignone2007MNRAS.378.1118M}.

Once a sufficiently precise solution for $W'$ is found, the primitive variables are given by \cref{eq: inversion rho}, $p(\chi, \rho)$ supplied by the EoS, and by
\begin{equation}
  \mathbf{u}(W') = \frac{\mathbf{m}}{(W' + D)^2 - \abs{\mathbf{m}}^2},
\end{equation}
\begin{table}
  \caption{
    Thermodynamical quantities for different EoS implemented in \cronos (see also \cref{sec: equations}) necessary for the variable inversion \citep[compiled from][]{Mignone2007MNRAS.378.1118M}.
    \label{tab: inversion thermodynamic quantities}
  }
  \centering
  \begin{tabular}{cccc}
    \hline\hline
              & $p(\chi, \rho)$                                                                               & $\partial_\chi p$                                     & $\partial_\rho p$                           \\
    \hline
    Ideal EoS & $\frac{\Gamma -1 }{\Gamma} \chi$                                                              & $\frac{\Gamma -1 }{\Gamma}$                           & 0                                           \\
    TM EoS    & $\frac{2 \chi (\chi + 2 \rho)}{5 (\chi + \rho) + \sqrt{9 \chi^2 + 18 \rho \chi + 25 \rho^2}}$ & $\frac{2 \chi + 2 \rho - 5 p}{5 \rho + 5 \chi - 8 p}$ & $\frac{2 \chi - 5 p}{5\rho + 5 \chi - 8 p}$ \\
    \hline
  \end{tabular}\end{table}

\subsubsection{Admissible conserved states}
If a given conserved state is unphysical, the variable inversion will fail and produce unphysical primitive states, e.g. with negative density or pressure, superluminal speeds, etc., or not converge at all.
To avoid related problems, we check the admissibility of a given conserved state $U \in \mathcal{G}_1$ before every inversion, relying on the set of physically-admissible states formulated by \citet{Wu2015JCoPh.298..539W}
\begin{equation}
  \mathcal{G}_1 = \lbrace {\bf U} \; \vert \; D > 0, \; q({\bf U}) \coloneqq \tau + D - \sqrt{D^2 + \abs{{\bf m}}^2} > 0 \rbrace
  \label{eq: admissible states condition}
  .
\end{equation}
If the constraint is violated ($U \notin \mathcal{G}_1$), the user is informed and the simulation is aborted by default.
Alternatively, the user can choose to overwrite the states in the concerned cells with new ones by supplying user-defined routines, e.g. setting a lower limit for the pressure and computing other quantities accordingly.

\subsubsection{Entropy inversion}
In situations where the thermal energy is negligible compared to the kinetic energy, simple discretization errors can lead to unphysical states ($\mathbf{U} \notin \mathcal{G}_1$) for which the variable inversion will fail.
To avoid related problems, we offer the user the possibility to solve an additional equation for the conservation of entropy in smooth flows
\begin{equation}
  \partial_t \Sigma + \nabla \cdot \left( \Sigma \mathbf{v}  \right) = 0, \label{eq: entropy conservation}
\end{equation}
next to the hydrodynamic equations \citep[in analogy to our procedure for classical HD and MHD][]{Kissmann2018ApJS..236...53K}, with $\Sigma = \sigma D$ and the entropy \citep{Mignone2007MNRAS.378.1118M}
\begin{equation}
  \sigma =
  \begin{cases}
    \frac{p}{\rho^\Gamma}            & \mathrm{ideal \, EoS} \\
    \frac{p}{\rho^{5/3}} (h -\Theta) & \mathrm{TM \, EoS}
    .
  \end{cases}
  \label{eq: entropy eos}
\end{equation}
Numerically, the equation is treated in analogy to passive tracer fields (see \cref{sec: tracer}), which guarantees to maintain positive values for $\Sigma$ using the implemented solvers.
Although the conservation of energy might be slightly violated by employing $\Sigma$ instead of $\tau$ in the variable inversion, the resulting scheme is much more stable, which can be necessary for certain applications.

When using $\Sigma$, the inversion is performed by numerically solving
\begin{equation}
  g(\rho) = \rho \, \glorentz( h(\rho) ) - D = 0 \label{eq: entropy inversion}
  ,
\end{equation}
where
\begin{eqnarray}
  \glorentz(h)
  = \sqrt{1 + \frac{\abs{\mathbf{m}}^2}{D^2 h^2}}
\label{eq: lorentz factor entropy inversion}
\end{eqnarray}
and
\begin{equation}
  h^2(\rho) =
  \begin{cases}
    \left(1 + \frac{\Gamma}{\Gamma - 1} \sigma \rho^{\Gamma - 1} \right)^2  & \mathrm{ideal \, EoS} \\
    \frac{\left( 1 + 4 \rho^{2/3} \sigma\right)^2}{1 + 3 \rho^{2/3} \sigma} & \mathrm{TM \, EoS}
  \end{cases}
  .
\end{equation}
Although a derivative of \cref{eq: entropy inversion} can be easily found and implemented for a Newton-Raphson scheme to solve the equation, we use a hybrid, bracketing, root-finding method, that is Brent's method \citep{brent1972algorithms}, instead.
For our use cases, we found a comparable performance for both solvers, while the latter one is guaranteed to converge in any case.
As initial bracket for the scheme we use $\rho_\mathrm{max} = D$ and $\rho_\mathrm{min} = D^2 / \sqrt{D^2 + \abs{\mathbf{m}}^2}$.
These limits are obtained from $\rho = D/\glorentz(h)$ using the asymptotic values of the specific enthalpy $ 1 < h < \infty$ in the ultra-cold and ultra-hot limit, respectively.

Once a sufficiently accurate value for $\rho$ is found, the inversion is completed by computing the fluid velocity from $\mathbf{u} = \mathbf{m} / D h(\rho)$ and the pressure according to
\begin{equation}
  p =
  \begin{cases}
    \sigma \rho^\Gamma                                                             & \mathrm{ideal \, EoS} \\
    \frac 1 2 \rho \left( h(\rho) - \sqrt{h^2(\rho) - 4 \sigma \rho^{2/3}} \right) & \mathrm{TM \, EoS}.
  \end{cases}
\end{equation}

The conservation of entropy, however, only applies for smooth, non-compressive flows.
To identify those cells for which this is the case and the variable inversion can be feasibly performed using \cref{eq: entropy inversion}, we flag cells according to the following criteria \citep[similar to][]{Balsara1999JCoPh.148..133B, Pluto}
\begin{eqnarray}
  \mathcal{F}^\Sigma_I =
\left( \norm{\nabla p_I} \max_d \Delta L^d_I < \alpha_1 \min_{I, N(I)} p \right)  \vee\\
  \left( - \alpha_2 \max_{I, N(I)} c_s < \left( \nabla \cdot \mathbf{v} \right) \min_d \Delta L^d_I \right) \nonumber
.
\end{eqnarray}
The occurring gradient and divergence are approximated using central, finite-differences and $c_s$ denotes the local sound speed computed according to \cref{eq: sound speed}.
We typically use values of $\alpha_1 = 0.05$ and $\alpha_2 = 0$.
Further, if a given cell $I$ does not qualify for the entropy-inversion ($\mathcal{F}^\Sigma_I = 0$), we also disable it for its direct neighbours for safety ($\mathcal{F}^\Sigma_J = 0$ for $J\in N(I)$).

Wherever $\mathcal{F}^\Sigma_I = 1$, we perform the variable inversion using \cref{eq: entropy inversion}.
Otherwise, \cref{eq: inversion equation} is used and the corresponding value of $\sigma$ is recomputed from the resulting state.
The described procedure can be enabled or disabled by a compile-time switch set by the user. 
\section{Validation} \label{sec: validation}
In the following sections, we present the results from a variety of numerical tests to verify our implementations within the \cronos code.

\subsection{1D Riemann problems}\label{sec: 1d riemann tests}
\begin{table}
  \caption{Parameters for the Riemann problems according to \cref{eq: validation Riemann problem} solved in \cref{sec: 1d riemann tests} \citep[taken from][]{Mignone2005MNRAS.364..126M} together with the $L_1$ errors of the numerical solutions for the density and the pressure, respectively.
  }\label{tab: 1d riemann parameters}
  \centering
  \small
  \begin{tabular}{ccccc}
    \hline\hline
            & P1                            & P2                              & P3                              & P4                                \\
    \hline
    $\vV_L$ & $\left( 1, 0.9, 1 \right)$ & $\left( 1, -0.6, 10 \right)$ & $\left( 10, 0, 40/3 \right)$ & $\left( 1, 0, 10^3 \right)$    \\
    $\vV_R$ & $\left( 1, 0, 10 \right)$  & $\left( 10, 0.5, 20 \right)$ & $\left( 1, 0, 2/3 \times 10^{-6} \right)$     & $\left( 1, 0, 10^{-2} \right)$ \\
    \hline
$L_1^\rho$ & $4.05 \times 10^{-2}$ & $5.62 \times 10^{-2}$ &  $5.94 \times 10^{-2}$ & $1.62 \times 10^{-1}$ \\
    $L_1^p$ & $6.88 \times 10^{-2}$ & $1.14 \times 10^{-1}$ & $3.23 \times 10^{-2}$ & $2.94$ \\
    \hline
  \end{tabular}
\end{table}\begin{figure*}
  \centering
  \includegraphics[width=\linewidth]{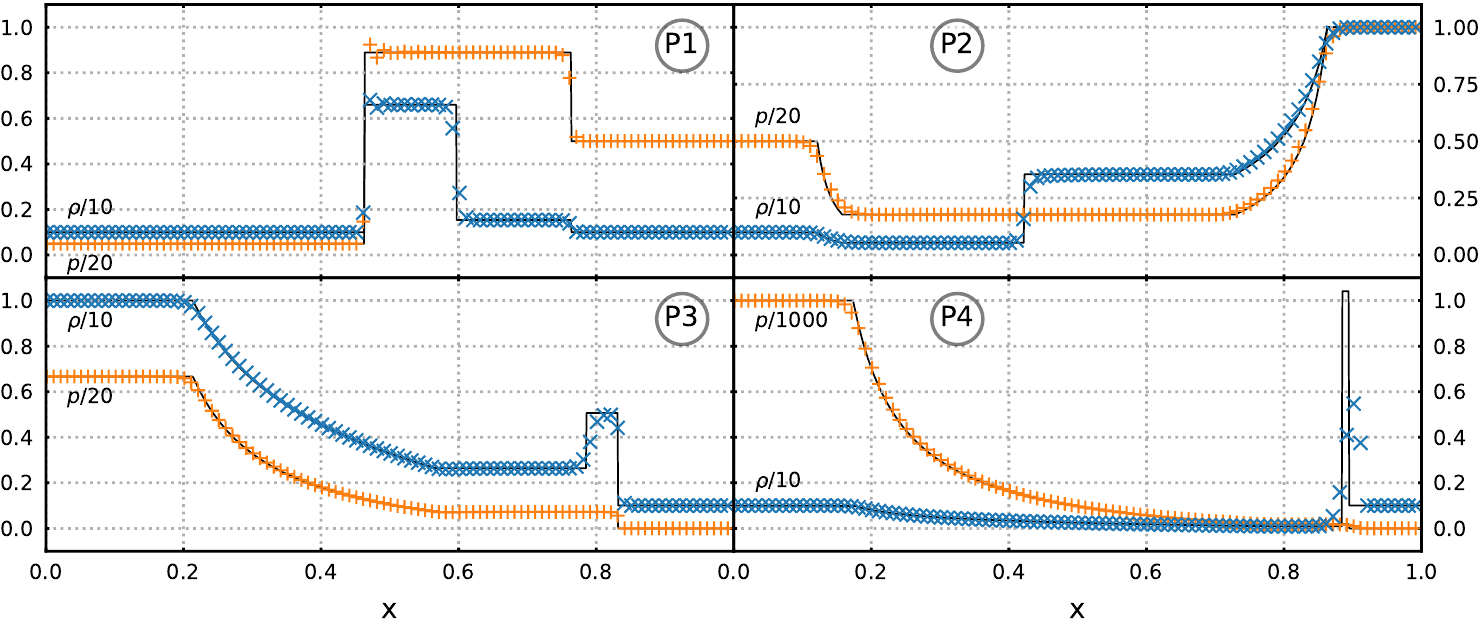}
  \caption{
    Numerical results for one-dimensional Riemann problems at time $t=0.4$.
    The density (blue crosses), and pressure (orange plusses) are shown.
    The underlying solid black line corresponds to the exact solution.
    The numerical solutions have been obtained using $400$ gridpoints, the minmod slope limiter, the \hllc solver, and CFL$=0.4$.
    To improve visibility, we show only every fourth point of the numerical results and scale the different quantities as annotated.
    The parametrisations of the Riemann problem for the numerical test can be found in \cref{tab: 1d riemann parameters}.
  } \label{fig: validation 1D riemann problem}
\end{figure*}One-dimensional Riemann problems are particularly useful numerical tests for hydrodynamic codes, since they can be solved by semi-analytical methods to arbitrary precision \citep[][]{Marti1994JFM...258..317M}.
The Riemann problems for the following tests are posed in the general form
\begin{equation}
  \left( \rho, v^x, p \right) = \begin{cases}
    \primitive_L & \text{for} \quad x < 0.5                                        \\
    \primitive_R & \text{for} \quad x > 0.5 \label{eq: validation Riemann problem}
  \end{cases}
  .
\end{equation}
For comparison we considered the same set of Riemann problems as in \citet{Mignone2005MNRAS.364..126M}, summarised in \cref{tab: 1d riemann parameters} with zero fluid velocity tangential to the discontinuity and an ideal EoS with $\Gamma = \frac{5}{3}$.
All numerical tests were performed using extrapolating boundary conditions.
\\
In \cref{fig: validation 1D riemann problem}, we present the numerical evolution of the considered Riemann problems together with their exact solutions \footnote{See \href{https://github.com/cxkoda/srrp}{github.com/cxkoda/srrp} for an implementation of an exact Riemann solver for relativistic hydrodynamics in \textsc{Python}.}.
In P1 and P2, two shock waves and two rarefaction waves are formed, respectively, which are in good agreement with the exact solutions.
The tests P3 and P4 are characterised by large differences in pressure between the initial states and evolve into a left-going rarefaction and a right-going shock wave, forming dense shells of material.
These shells are also known as relativistic blast waves and pose notoriously difficult problems for numerical schemes.
For P4, the numerical solution yields a shock position and a compression ratio that differ slightly from their exact locations and values.
This is caused by an insufficient resolution of the blast wave in combination with the spatial reconstruction effectively reducing to first order in the vicinity of shocks.
Overall, the numerical solutions obtained using \cronos are in good agreement with the exact, semi-analytic solutions, which validates the implemented solvers.

\subsection{1D Riemann problems with non-zero tangential velocity}
\begin{figure}
  \centering
  \includegraphics[width=\linewidth]{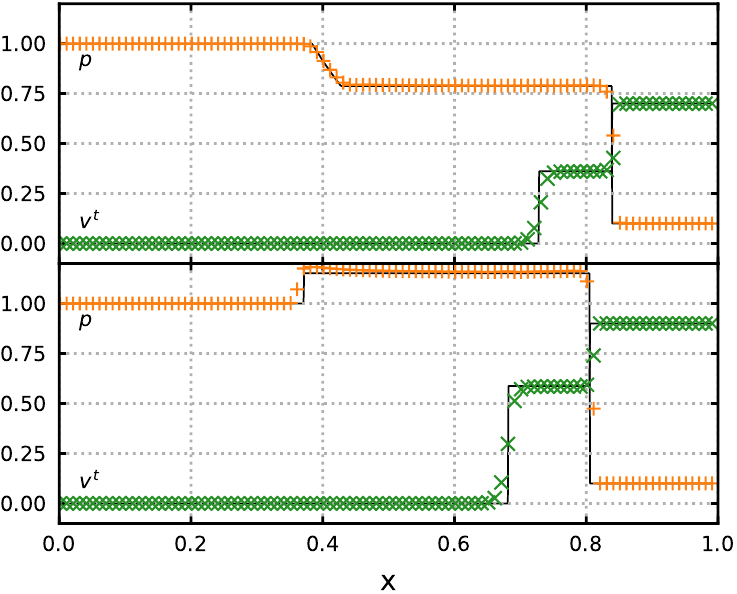}
  \caption{
    Same as in \cref{fig: validation 1D riemann problem}, but for Riemann problems involving non-zero tangential velocities in the right state with $v^t_R=0.7$ (top) and $v^t_R=0.9$ (bottom) as parameterised in \cref{eq: validation 1D riemann problem nonzero}.
    The solutions for the tangential velocity (green crosses) and the pressure (orange plusses) are shown.
  } \label{fig: validation 1D riemann problem nonzero}
\end{figure}

While velocities tangential to the initial discontinuity $v^t$ do not affect the evolution of Riemann problems in classical hydrodynamics, they do have a severe impact on the emerging wavefan in relativistic hydrodynamics because of the non-linear coupling introduced by the Lorentz factor.
We demonstrated that these effects are also correctly handled by our implementations by solving the Riemann problem
\begin{equation}
  \left( \rho, v^x, v^t, p \right) = \begin{cases}
    \left( 1, 0.5, 0, 1 \right)         & \text{for} \quad x < 0.5                                                   \\
    \left( 0.125, 0, v^t_R, 0.1 \right) & \text{for} \quad x > 0.5 \label{eq: validation 1D riemann problem nonzero}
  \end{cases}
\end{equation}
for different values of the right-sided, tangential velocity $v^t_R$ in analogy to the ones in \cref{sec: 1d riemann tests}.
Exact semi-analytic solutions with non-zero tangential velocities are again available in the literature \citep[see][]{Pons2000JFM...422..125P,Rezzolla2003JFM...479..199R}, which were used to produce the reference solutions shown in \cref{fig: validation 1D riemann problem nonzero}.

We investigated two scenarios with $v^t_R=0.7$ and $v^t_R=0.9$ and show the numerical solutions in \cref{fig: validation 1D riemann problem nonzero}.
The numerical results agree well with the exact solutions and show the transition from a rarefaction-shock to a shock-shock pattern expected at $v^t_R \sim 0.85$ \citep[][]{Rezzolla2003JFM...479..199R}.
The numerical $L_1$ errors are $L_1^\rho = 6.20 \times 10^{-3}; 9.76 \times 10^{-3}$ for the density and $L_1^p = 6.61 \times 10^{-3}; 9.04 \times 10^{-3} $ for the pressure, corresponding to $v^t_R=0.7; 0.9$, respectively.

\subsection{2D Riemann problem}
\begin{figure}
  \centering
  \includegraphics[width=\linewidth]{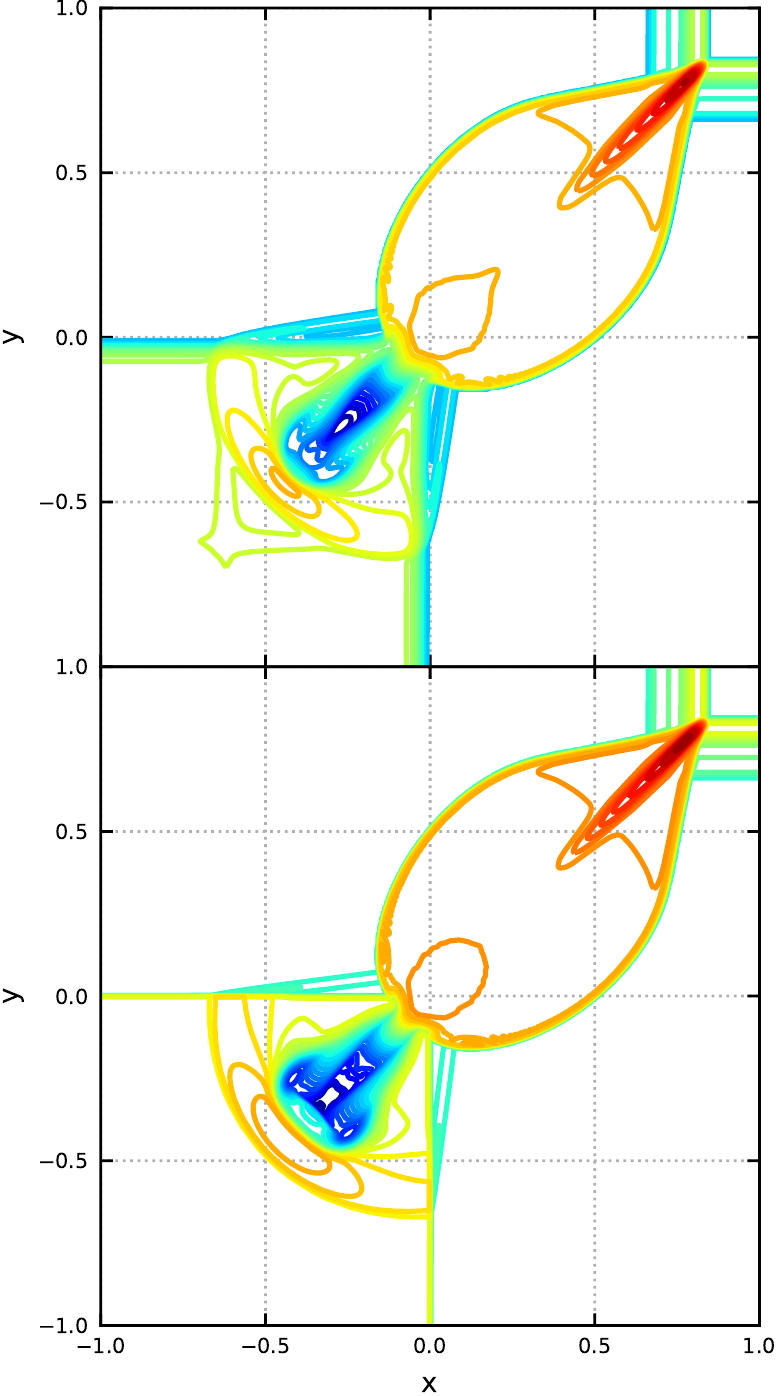}
  \caption{
    Logarithmically equidistant contours of the rest-mass density for numerical solutions of the two-dimensional Riemann problem given in \cref{eq: validation 2D riemann problem} at time $t = 0.8$.
    The results using the \hll solver (top) and \hllc solver (bottom) are shown, respectively.
    The solutions were obtained on a grid with $400^2$ cells using a CFL number of $0.4$
  } \label{fig: validation 2D riemann problem}
\end{figure}
After the successful tests in one dimension, we now move to tests in higher dimensions.
First, we consider a two-dimensional Riemann problem, which is posed by the interaction of four different initial states; one in each quadrant of the two dimensional plane.
For the sake of comparison, we solved the same problem as investigated by \citet{DelZanna2002A&A...390.1177D} that is parameterised as
\begin{equation}
  \left( \rho, v^x, v^y, p \right) = \begin{cases}
    \left( 0.1, 0, 0,0.01 \right)  & \text{for} \quad x > 0, \; y > 0 \\
    \left( 0.1, 0.99, 0, 1 \right) & \text{for} \quad x < 0, \; y > 0 \\
    \left( 0.5, 0, 0,1 \right)     & \text{for} \quad x < 0, \; y < 0 \\
    \left( 0.1, 0, 0.99,1 \right)  & \text{for} \quad x > 0, \; y < 0.
  \end{cases} \label{eq: validation 2D riemann problem}
\end{equation}
Again, we use extrapolating boundary conditions, the minmod slope limiter and an ideal EoS with $\Gamma = \frac{5}{3}$.\\
In \cref{fig: validation 2D riemann problem}, we present the numerical results using the \hll and \hllc solver, respectively.
The initial discontinuities separating the two top states and the two right states evolve into curved shock waves that enclosing a drop-shaped region in the top-right quadrant.
The lower-left region reamins confined by two contact discontinuities and forms a jet-like structure.
While the contact discontinuities remain sharp and steady using the \hllc solver, they degenerate into spurious waves using the \hll solver because of the increased numerical dissipation of the latter, demonstrating the better performance of the \hllc solver.
Our numerical results agree well with the ones in the literature \citep[e.g.][]{DelZanna2002A&A...390.1177D, Mignone2005MNRAS.364..126M}.
Having verified the implementation on Cartesian grids, we move to curvilinear coordinates in the next test.

\subsection{Axisymmetric relativistic jet}\label{sec: relativistic jet}
\begin{figure}
  \centering
  \includegraphics[width=\linewidth]{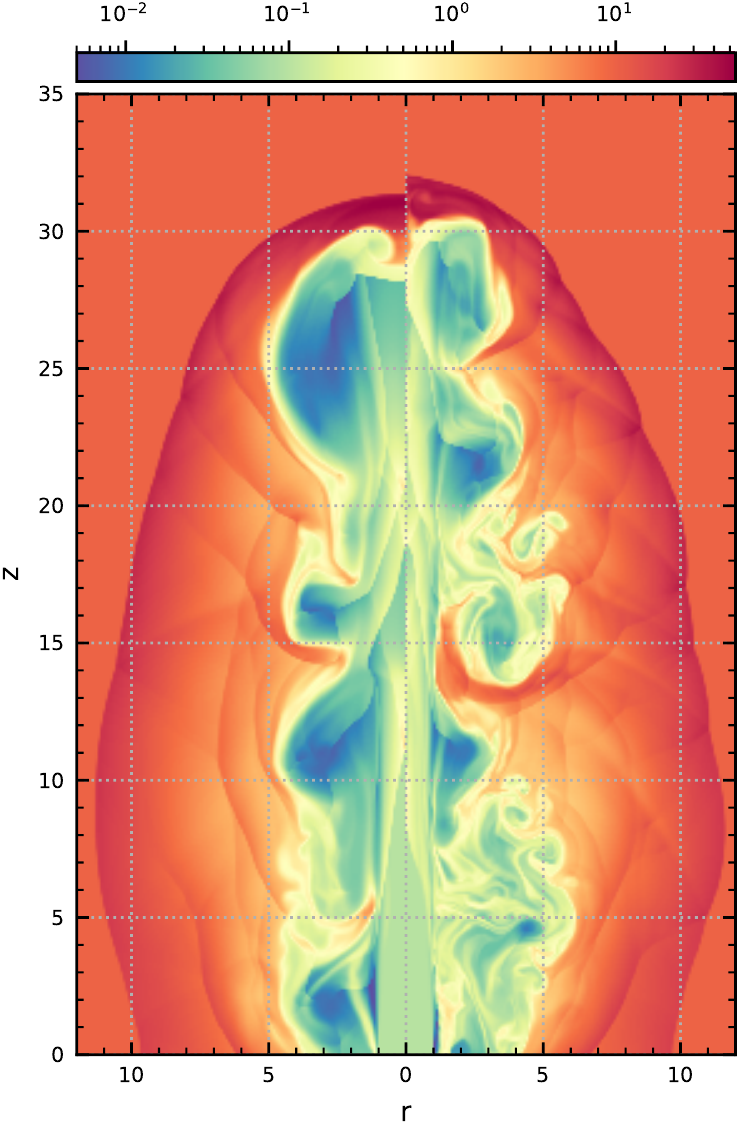}
  \caption{
    Rest-mass density for the relativistic jet given in \cref{sec: relativistic jet} at time $t=80$.
    The results using the \hll solver (left half) and \hllc solver (right half) are shown, respectively.
    The numerical solutions were obtained on a grid with $512 \times 1536$ cells in cylindrical axisymmetry using a CFL number of $0.4$.
  } \label{fig: 2D relativistic jet}
\end{figure}In this numerical test we demonstrate the implementation of geometric source terms arising in non-Cartesian coordinate systems (see \cref{sec: coordinate systems}).
For this, we simulated the evolution of an axisymmetric relativistic jet in 2D cylindrical geometry.
The jet is set up by initialising the computational volume with an ambient, background medium (respective quantities are denoted with an index $m$) and injecting the supersonic beam material (denoted with an index $b$) by imposing constant boundary conditions at $z=0$ for radii $r<1$.
At the symmetry axis of the jet $r = 0$, we impose reflecting boundary conditions and outflowing boundary conditions everywhere else.
The computational domain covers the region $0 \leq r \leq 12$ and $0 \leq z \leq 35$.
For the sake of comparison, we have adopted the same set of parameters as \citet{DelZanna2002A&A...390.1177D}, which are given by the beam density $\rho_b = 0.1$, the beam speed $v_b^z = 0.99$, the ambient medium density $\rho_m = 10$, matched pressures $p_b = p_m = 10^{-2}$, and an ideal EoS with $\Gamma = \frac{5}{3}$.
The ambient medium is at rest $v_m=0$ and the beam has no radial velocity $v_b^r = 0$.

The results at time $t=80$ are shown in \cref{fig: 2D relativistic jet} and exhibit typical features of relativistic jets, such as the bow-like shock of the ambient medium, a hull of shocked material, a central relativistic stream and turbulence excited by the Kelvin-Helmholtz instability due to the high velocity shear.
Simulations using the \hllc solver show more small-scale fluctuations than those using the \hll solver because of the latter's higher numerical dissipation.
This again demonstrates the superior performance of the \hllc solver.
The results are in good qualitative agreement with the numerical solutions obtained by \citet{DelZanna2002A&A...390.1177D} and \citet{Mignone2005MNRAS.364..126M}.
The slight extension of the head using the \hllc solver is potentially caused by the carbuncle-problem, which usually manifests itself as an extended nose on the jet-axis \citep{Quirk1994IJNMF..18..555Q} and has also been seen in other implementations of this test \citep[e.g. see][]{Lamberts2013A&A...560A..79L}.
\\
In this test, we exclusively considered a single ideal EoS.
The evolution of relativistic jets, however, is heavily affected by the particular choice of the EoS, which is exploited in the next tests.

\subsection{Equation of state}\label{sec: validation eos}
\begin{figure}
  \centering
  \includegraphics[width=\linewidth]{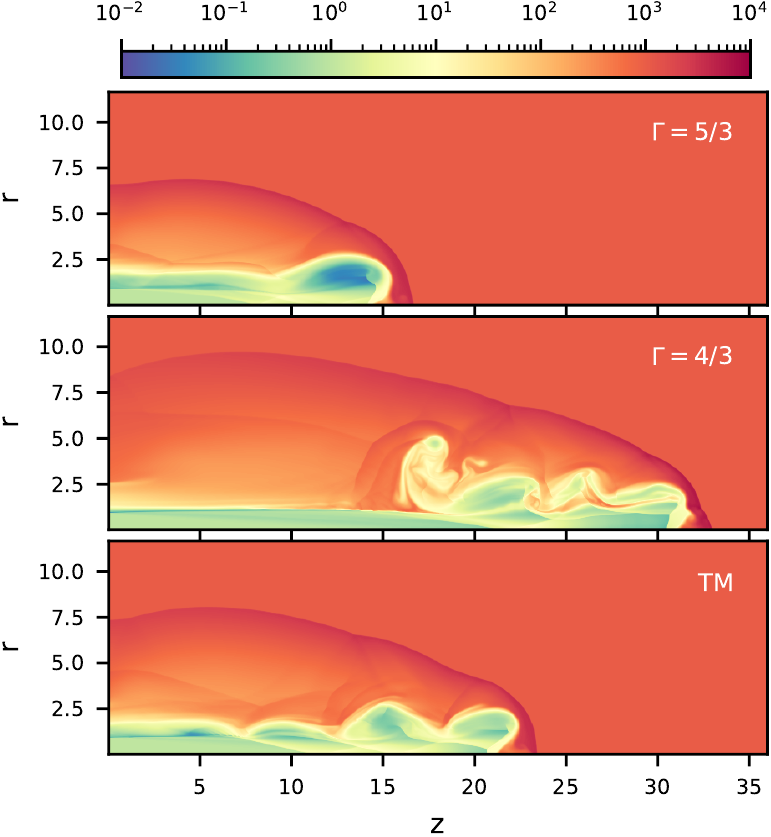}
  \caption{
  Rest-mass densities for the jets simulated with different equations of state (as annotated) at $t = 60$.
  The computations were performed on a grid with $512 \times 1536$ cells in cylindrical symmetry using the \hllc solver and $\mathrm{CFL}=0.4$.
  The jets were set up using the parameters given in the text and \cref{tab: eos jet params}.
  \label{fig: eos jet}
  }
\end{figure}

\begin{table}
  \caption{
    Ambient medium densities and one-dimensional estimates for the jet speed according to \cref{eq: analytic jet speed} for the jets considered in \cref{sec: validation eos}.
    \label{tab: eos jet params}}
  \centering
  \begin{tabular}{cccc}
    \hline\hline
    & $\Gamma = \frac{5}{3}$ & $\Gamma = \frac{4}{3}$ & TM \\
    \hline
    $\rho_m$ & 27.7424 & 2.19394 & 10.3337\\
    $V_j$ & 0.30354 & 0.57883 & 0.39243\\
    \hline
  \end{tabular}
\end{table}

Here we verify the different EoS implemented in \cronos (see also \cref{sec: equations}).
For this, we have again simulated the propagation of relativistic jets in analogy to the setup in \cref{sec: relativistic jet}.
For a given equation of state, the properties of a jet are determined by three parameters: the density ratio $\eta = \rho_b / \rho_m$, the beam's Lorentz factor $\glorentz_b$ and its Mach number $M_b = v_b / c_s$.
Following the tests performed in \citet{Mignone2007ApJS..170..228M}, we have adopted the same jet parameters $\eta = 10^{-3}, \glorentz_b = 10, M_b = 1.77$ and investigate the jet evolution under three different equations of state: two ideal ones with $\Gamma = \frac{5}{3}$ and $\Gamma = \frac{4}{3}$ and the TM EoS.
We further chose the pressures $p_b = p_m = 10^{-2}$ resulting in the ambient medium densities $\rho_m$ listed in \cref{tab: eos jet params}, which were used together with $\rho_b = \eta \, \rho_m$ and $v_b = 0.995$ to specify the setup for the simulation.

\begin{figure}
  \centering
  \includegraphics[width=\linewidth]{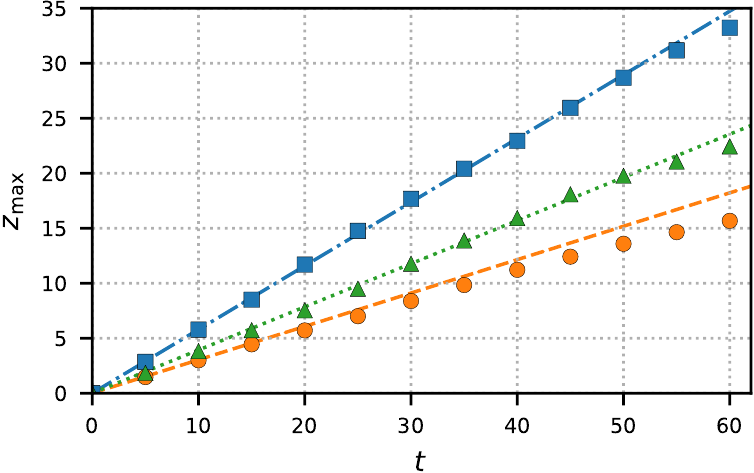}
  \caption{Position of the head of the jet over time for different equations of state: $\Gamma = \frac{5}{3}$ (orange, circles, dashed),  $\Gamma = \frac{4}{3}$  (blue, squares, dash-dotted), TM-EoS (green, triangles, dotted).
  Markers correspond to numerical results and lines to one-dimensional estimates given by \cref{eq: analytic jet speed}.
  The simulated jets correspond to the ones depicted in \cref{fig: eos jet}.
  \label{fig: jet head}
  }
\end{figure}In \cref{fig: eos jet} we show the resulting densities for the simulated jets at $t=60$.
Owing to the different EoSs, the head of the jets propagate at different speeds along the z-axis, which is visible from the different jet extents in the simulation.
\\
In one dimension, the propagation of relativistic jets was investigated by \citet{Marti1997ApJ...479..151M} who derived the following analytical expression for the speed of the head of a jet
\begin{equation}
  V_\mathrm{j} =\frac{\glorentz_b \sqrt{\eta \, h_b/h_m}}{1 + \glorentz_b \sqrt{\eta \, h_b/h_m}}
  \label{eq: analytic jet speed}
  .
\end{equation}
The values corresponding to our setups can again be found in \cref{tab: eos jet params}, which serve as a test for our implementations.
In \cref{fig: jet head} we show the locations of the jet-heads in our numerical simulations, which are in good agreement with the analytical estimates from \cref{eq: analytic jet speed}.
As expected from similar simulations performed by \citet{Mignone2007ApJS..170..228M}, the simulated jets become slower and start to deviate from the linear propagation in one dimension for later times, which follows from the two-dimensional nature of the numerical tests.

\subsection{Spatial order}\label{sec: order}
Finally, we verify the order of our scheme using similar tests than we performed for classical hydrodynamics with \cronos \citep[see][]{Ryu1994ApJ...422..269R, Kissmann2018ApJS..236...53K}.
These tests investigate the decay of two-dimensional standing sound waves by the numerical dissipation of the scheme.
\\
In analogy to classical hydrodynamics, also in special relativity, acoustic waves can be excited by adding velocity perturbations to a constant background state, owing to the similar structure of the eigenvectors of the linearised equations \citep[see][]{Rezzolla2013}.
We initialised sinusoidal sound waves using
\begin{equation}
  \delta u^x = \delta u^y = \delta_\mathrm{amp} \sin(k_x x + k_y y)
  ,
\end{equation}
with the perturbation amplitude $\delta_\mathrm{amp} = 10^{-5}$, the wavenumbers $k_x = k_y = \frac{2 \pi}{L}$ and the side length of the computational domain $L=1$.
This produces two acoustic waves that propagate with velocities $\lambda_\pm(\primitive_0)$ into opposite directions, which can be computed from \cref{eq: characteristic velocities} for a given constant background state $\primitive_0$.
For the purpose of this test, we chose the background density $\rho_0 = 1$ and the four-velocity $u^x_0=u^y_0=0, u^z=\mathrm{const.}$ to ensure that both waves travel at the same speed, yielding
\begin{equation}
  \lambda_\pm = \pm c_s \sqrt{\frac{1 - (v^z_0)^2}{1-  (v^z_0)^2 c_s^2}}
  ,
\end{equation}
where $v^z_0 = \frac{u^z_0}{\sqrt{1 + (u^z_0)^2}}$ is the three-speed in $z$-direction and $c_s$ the sound speed of the background medium, respectively.

In a periodic domain, this produces a standing wave that oscillates with frequency $f = \frac{\sqrt{2}}{L \, \lambda_+}$.
To introduce relativistic effects, we set the background speed to a relativistic value of $u^z_0 = \frac{2}{\sqrt{3}}$.
Choosing $c_s = \frac{1}{2}$, which implies $p_0 = \frac{3}{4 \rho_0}$ for an ideal EoS with $\Gamma = \frac{4}{3}$, therefore leads to one oscillation in two units of time.
By simulating until $t=20$, we thus captured 10 full oscillations.

From the simulations, we retrieved the spatial root-mean-squared average of the velocity perturbation that oscillates with the double frequency.
Due to the presence of numerical viscosity, its amplitude decreases exponentially $\propto \exp(- \Gamma_\mathrm{decay}\,t)$.
The decay rate $\Gamma_\mathrm{decay}$ was determined from the simulation results as a function of spatial resolution and is shown in \cref{sec: order}.
The second-order nature of our scheme is obvious from $\Gamma_\mathrm{decay} \propto N^{-2}$.
As expected the \hllc solver introduces less numerical viscosity as compared to the \hll solver.
In both cases we used the minmod slope limiter and CFL$=0.4$.
\begin{figure}
  \centering
  \includegraphics[width=\linewidth]{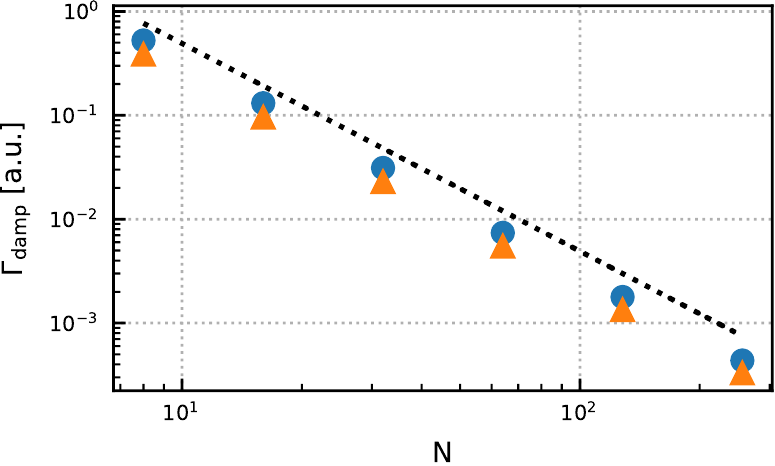}
  \caption{
    Damping rate $\Gamma_\mathrm{decay}$ for the decaying, sound-waves test (see \cref{sec: order}) as a function of the number of grid cells $N$ using the \hll (blue circles) and the \hllc solver (orange triangles), respectively.
    The black dotted line indicates a $N^{-2}$ dependence to guide the eye.
    }
  \label{fig: order}
\end{figure}%
\section{Parallel Performance} \label{sec: performance}
\subsection{Weak scaling}
\begin{figure}
    \centering
    \includegraphics[width=\linewidth]{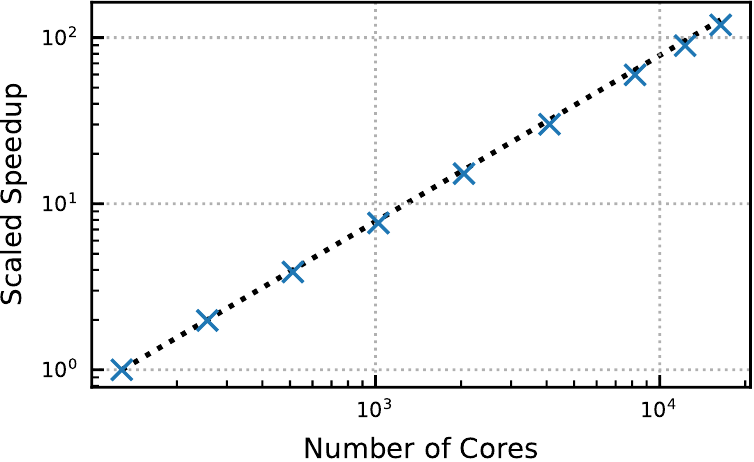}
    \caption{Scaled speedup of simulations performed using \cronos as a function of the number of computation cores in a weak-scaling test.
        The dotted line indicates perfect scaling.
    }
    \label{fig: scaling}
\end{figure}

The relativistic extension of \cronos has been successfully employed in various computational environments using up to $\sim 10^4$ computing cores.
The code is parallelised using a halo-communication approach that was realised employing the Message Passing Interface (MPI) library.

To test the parallel efficiency of our code, we performed a weak-scaling test on the Joliot-Curie Rome HPC-environment with AMD Rome (Epyc) 7H12 2.6GHz bi-processor compute nodes.
As a test setup, we considered the colliding winds of two spherically symmetric, static sources similar to the simulation of gamma-ray binaries in \citet{Huber2020}.
Initially, the system was divided into two regions, each filled by one of the winds, using an approximate curve for the contact discontinuity in the system.
The simulation was run sufficiently long to obtain a representative average of the time step duration.
While the domain size was increased with each test, the workload handled by each computation core was kept constant at $64^3$ cells.
The results are shown in \cref{fig: scaling}.
As expected, the parallel efficiency drops with an increasing number of cores but remains above $93\%$ for more than $16$k cores.

\section{Summary}\label{sec: summary}
We have presented and verified an extension to the \cronos code treating special relativistic hydrodynamics for astrophysical applications.
The code is based on a semi-discrete finite-volume framework, ensuring the conservation of relevant quantities to efficiently capture shocks.
\cronos employs a second-order spatial reconstruction combined with a second-order Runge-Kutta time-integrator.
The extension was developed with a focus on robustness, providing the user with additional interfaces to prevent, detect and possibly counteract failures.

Our implementation was verified against a set of standard problems, for which either an analytical solution or established numerical solutions from the literature are available.
Parallel scaling for up to $\sim 10^4$ cores has been demonstrated.

As the rest of the \cronos framework, this extension was also developed with modularity in mind, making it easy to consider additional physics supplied by the user.
This was, for example, used to simulate an additional transport of energetic particles governed by a Fokker-Planck type equation alongside the fluid dynamics \citep[as demonstrated in][]{Huber2020, Huber2021}.

\begin{acknowledgements}
  We thankfully acknowledge the access to the research infrastructure of the Institute for Astro- and Particle Physics at the University of Innsbruck (Server Quanton AS-220tt-trt8n16-g11 x8), the LEO HPC infrastructure of the University of Innsbruck and PRACE resources, that were used throughout the development of the code.
  We acknowledge PRACE for granting us access to Joliot-Curie at GENCI@CEA, France.
  D.H. acknowledges financial support by the University of Innsbruck through the stipend grand 2020/2/MIP-15 titled "Doktoratsstipendium aus der Nachwuchsförderung".
  This research made use of Cronos \citep{Kissmann2018ApJS..236...53K}; GNU Scientific Library (GSL) \citep{galassi2018scientific}; matplotlib, a Python library for publication quality graphics \citep{Hunter2007}; Scipy \citep{2020SciPy-NMeth}; and NumPy \citep{2020NumPy-Array}.
  We thank the anonymous referee for the thoughtful comments and suggestions that allowed us to improve our manuscript.
\end{acknowledgements}

\bibliographystyle{aa} \bibliography{references}

\begin{appendix} \section{Geometrical source terms}\label{app: geom source terms}
\begin{table}
  \caption{
    Coordinate variables $x^i$, scale factors $h_i$ and geometrical source terms $G_{m^j}$ for the different coordinate systems implemented in \cronos.
    \label{tab: coordinates}
  }
  \centering
  \begin{tabular}{rccc}
    \hline\hline
                      & Cartesian   & Cylindrical                   & Spherical                                                                \\
    \hline
    $(x^1, x^2, x^3)$ & $(x, y, z)$ & $(r, \phi, z)$                & $(r, \theta, \phi)$                                                      \\
    $(h_1, h_2, h_3)$ & $(1,1,1)$   & $(1,r,1)$                     & $(1,r,r \sin \theta)$                                                    \\
    $G_{m^1}$         & 0           & $\frac{p + m^\phi v^\phi}{r}$ & $\frac{2 p + m^\theta v^\theta + m^\phi v^\phi}{r}$                      \\
    $G_{m^2}$         & 0           & $\frac{- m^\phi v^r}{r}$      & $\frac{- m^\theta v^r + \left(p + m^\phi v^\phi \right)\cot \theta}{r}$ \\
    $G_{m^3}$         & 0           & 0                             & $\frac{- m^\phi \left(v^r + v^\theta \cot \theta \right)}{r}$            \\
    \hline
  \end{tabular}\end{table}
The geometrical source terms $G_{m^j}$ arising in \cref{eq: momentum eq with geom source} can be obtained using the anholonomic connection symbol  $\Omega^{j}_{ik}$ in the given coordinate system as
\begin{eqnarray}
  G_{m^j} = - \Omega^{j}_{ik} T^{ik}
  ,
\end{eqnarray}
where $T^{ij} = m^i v^j + p\,  \delta^{ij}$ denotes the stress-tensor components.
The connection symbol corresponds to the anholonomic components of the Christoffel symbol of second kind $\Gamma^{j}_{ik}$ (also known as natural connection) in physical units and can be computed by \citep[see][]{Altman2011}
\begin{equation}
  \Omega^{j}_{ik}= \frac{h_j}{h_i h_k} \Gamma^{j}_{ik} - \frac{h_j}{h_k} \partial_k ( 1/ h_i) \, \delta^j_i
  .
\end{equation}
This yields the geometrical source terms
\begin{equation}
  G_{m^j}= - \sum_{i,k}  \frac{h_j}{h_i h_k} \Gamma^{j}_{ik} T^{ik} + \sum_i \frac{\partial_i h_j}{h_i h_j} T^{ij}
  .
\end{equation}
The Christoffel symbol can be readily obtained from the metric tensor $g_{ij} = h_i h_j \delta_{i,j}$ and its derivatives via
\begin{equation}
  \Gamma^a_{bc} = \frac{1}{2} g^{ai} \left(
  \partial_b g_{ci}
  + \partial_c g_{ib}
  - \partial_i g_{bc}
  \right)
  .
\end{equation}
For orthogonal coordinates \citep{Stone1992ApJS...80..753S} the only non-zero entries are given by
\begin{eqnarray}
  \Gamma^i_{ii} = \frac{\partial_i h_i }{h_i}; \,  \Gamma^i_{jj} = \frac{h_j}{h^2_i} \partial_i h_j ; \,  \Gamma^i_{ij} =\Gamma^i_{ji} = \frac{ \partial_j h_i}{h_i}
  .
\end{eqnarray}
The resulting source terms for the implemented coordinate systems are summarised in \cref{tab: coordinates}.
We note that the source terms arising in special relativistic hydrodynamics are formally identical to the ones in Newtonian hydrodynamics.
This is intuitive since both sets of equations are also formally identical except for different definitions of the conserved variables.
\\
A similar procedure can also be used to derive the source terms for co-rotating coordinate systems in special relativistic hydrodynamics.
This has been done in \citet{Huber2020} in the context of simulation of gamma-ray binaries and is also available in \cronos.

 \end{appendix}
\end{document}